# Laser induced ponderomotive convection in water

M.N. Shneider[1*] and V.V. Semak

[1]Mechanical and Aerospace Engineering Department, Princeton University, NJ 08544, USA

## ABSTRACT

A new mechanism for inducing convection during IR laser interaction with water or any absorbing liquid is described theoretically. The numerical simulations performed using the developed model show that the optical pressure and ponderomotive forces produces water flow in the direction of the laser beam propagation. In the later stage of interaction, when water temperature rises, the Archimedes force becomes first comparable and then dominant producing convection directed against the vector of gravitational acceleration (upward). The theoretical estimates and the numerical simulations predict fluid dynamics that is similar to the observed in the previous experiments.

## I. INTRODUCTION

Water is a fascinating and rather controversially discussed substance in condensed matter research [1]. It is well known that water is the natural substance that has the most known anomalous properties. It is also fair to admit that physics of water is ill understood and it is to large extent a mystery in spite the facts that two thirds of the planet is covered with water, that water makes our environment a livable place, that human body is 53% water by mass, and in spite that the mystery of life itself is closely related to the mystery of water since the DNA molecule has its 3D helical form only in the water environment.

One should wonder of how such ubiquitus yet of extreme practical importance substance recieved so little of intelectuall attention. A plasusible explanation is in the longtime trend of the science funding suppor being mainly offered to the research that promis either exotic outcome or very practical outcome with very high probabilty of success obtained very quickly. This leaves very little or no room for studies of "ubiquitus" substances and phenomena. However, in few instances demand for an exotic outcome requires study of the omnipresent water or any other fluid for that matter. An example of that is recent (within the past decade) raise of optofluidics, a realm of research and technology that deals with manipulation of molecules and nanoparticles suspended, typically, in water or in other fluid [2].

In the present work we describe research that also emerged from an applied research project with a goal of transmitting large power through water. In this project we have encountered need for a study of collective behavior of water molecules in high gradient fast varying electromagnetic fields. Thus, the niche of our research scale wise is on micrometer to millimeter scale covering gap between the nano-scale, as in case of optofluidics, and the macro-scale described by hydrodynamics.

Investigation of behavior of liquid dielectrics in electric fields has a long, however, not so rich history, which was started by Michael Faraday [3] in early 19 century. It is known that dielectric fluids in strong nonuniform electric field are influenced by electrostrictive ponderomotive force [4–6]. As a result, fluid tends to be set in motion and moves into the regions with the strongest field. It was shown that the shape of liquid droplets can be modified by the volumetric electrostrictive forces arising in the vicinity of inhomogeneous laser beam [7].

[*] m.n.shneider@gmail.com

Additionally, it was theoretically demonstrated that the electrostrictive forces induced by the laser in liquid could be the source of acoustic pulses [8,9]. In another example the volumetric ponderomotive force occurs when a high voltage is rapidly applied to a sharp needle-like electrode produce region with significant negative pressure leading to the rupture of liquid, i.e. to cavitation, that in turn facilitates development of the electric breakdown [10-14].In our previous experimental project one of the objectives was to create a vapor channel in water using high power fiber laser operating at wavelength of 1.064 μm. According to our theoretical estimates the absorption of laser power will produce increase of temperature and related surface evaporation that will be sufficient to create vapor cavity similarly to formation of keyhole in laser welding [15,16]. Contrary to the theoretical estimates, focused laser beam did not produce expected evaporation of water within applied power range of up to 7kW. Instead, we observed steady water flow at lower laser power that, with increase of laser power, was becoming unstable, turbulent, and, at higher powers, rather violent.

The schematic layout of the experiment is shown in the Figure 1. The example of the observed fluid dynamics is shown in the Figures 2 and 3 that represent sequence of frames of the shlieren video taken orthogonally to the fiber laser beam. The fiber laser beam propagation is vertical from top down in the Figures 2 and 3, and the upper horizontal edge of the frame is at the water surface. The fiber laser beam with approximate radius of 1.2cm on $1/e^2$ level of intensity was focused by a 20 cm focal length lens such that the focal plane was at the depth of 5cm (below the bottom of the frame). Thus, the radius of the laser beam at the water surface was approximately 3mm at the $1/e^2$ level of intensity. The shlieren video was recorded using HD Canon camcorder and the frames shown in the Figures 2 and 3 were within 2 seconds from the beginning of interaction of laser beam with water. The time between the shown sequent frames is approximately 0.1s. Unfortunately more acurate timing information can not be provided since we did not expect to observe fast dynamic phenomenon and, therefore, a slow framimng rate camcorder was used.

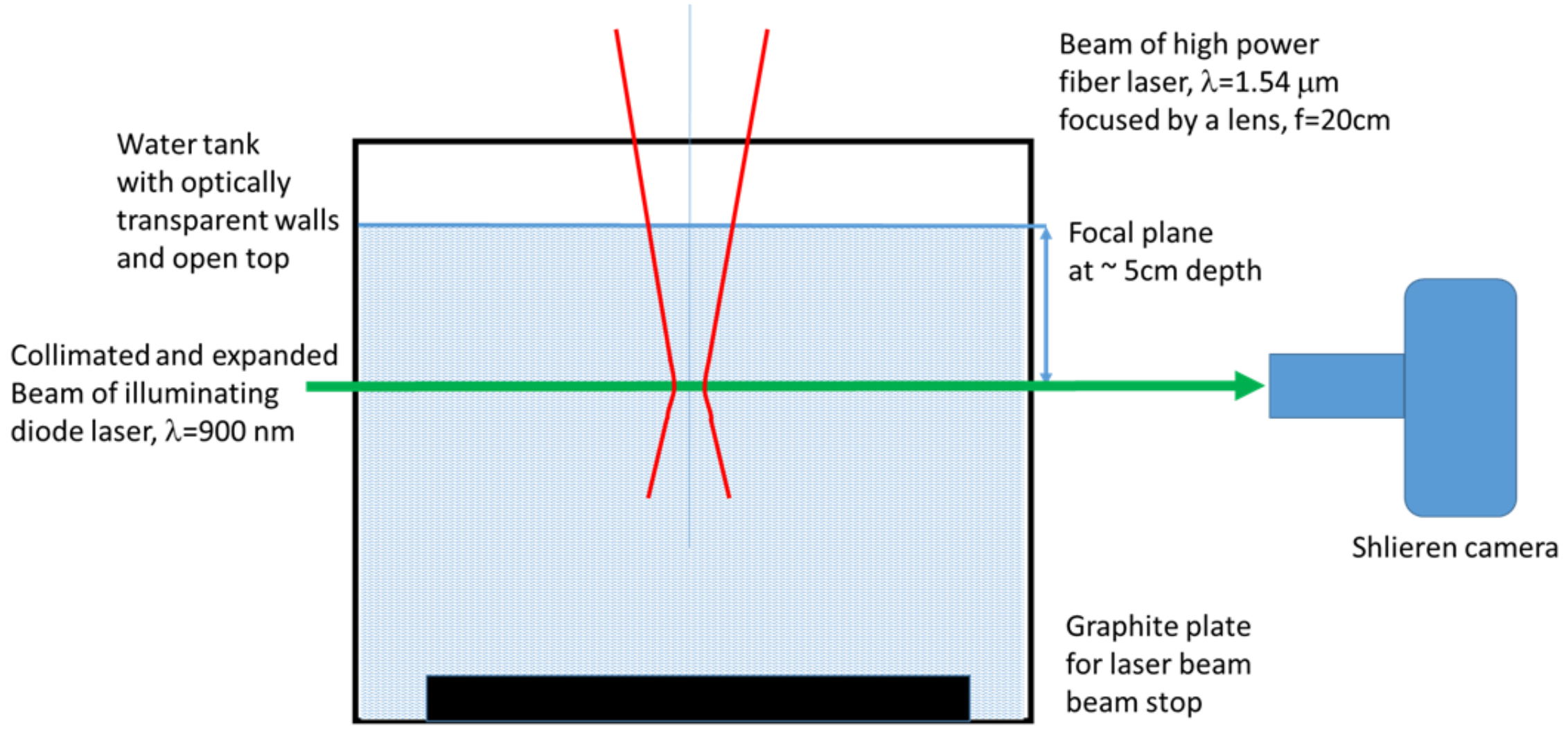

Fig. 1. Experiment schematics.

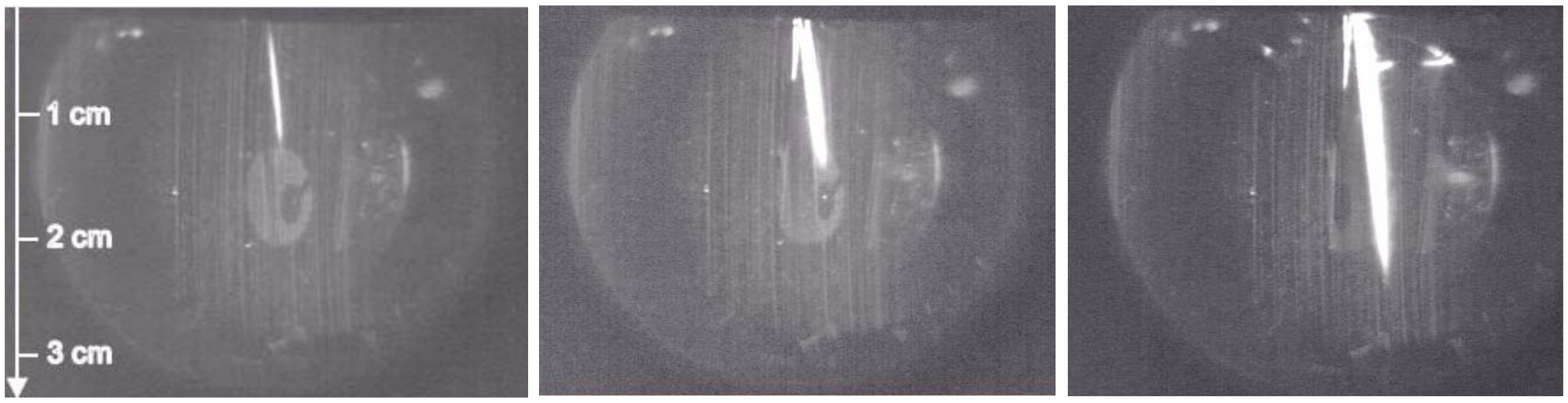

Fig. 2. $P_l$=150 Watt. Time between the frames $\Delta t \sim 0.1$ sec; Remains stable for a long time ~ seconds

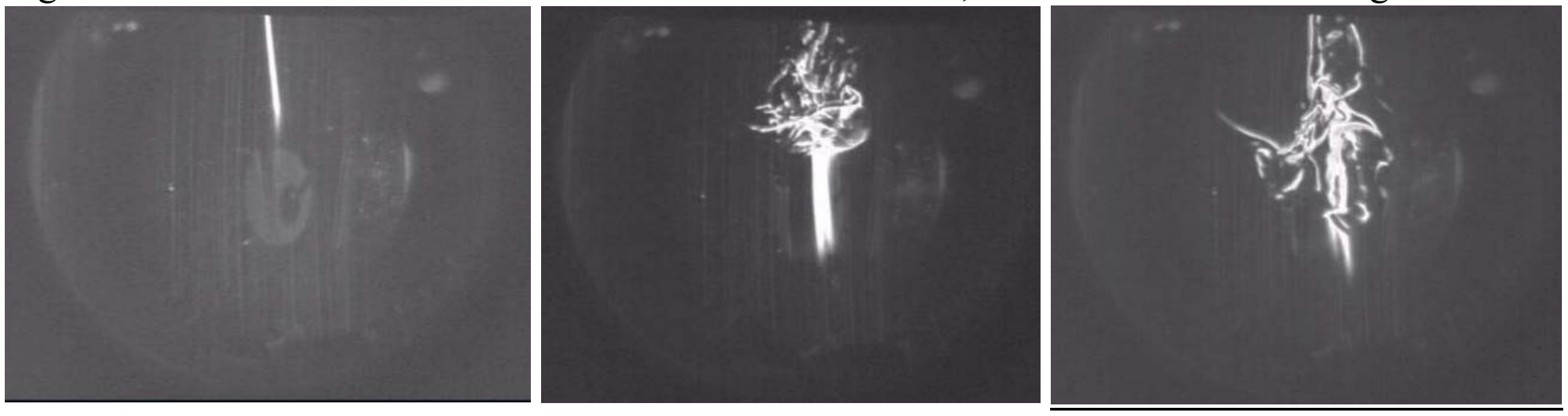

Fig. 3. $P_l$=380 Watt. Time between the frames $\Delta t \sim 0.1$ sec; Instability develops and turbulization starts in time less than 0.5 sec

The absorption of the laser beam produces increase of the water temperature and, therefore, a thermally induced upward convection is expected to develop in correlation with water heating. However, in experiments we observed significant delay of thermally induced convection. It is curious that the delay of thermal convection is independent of the laser power. In this work we show that this delay is produced by the optical pressure from the absorbed laser beam and by the volumetric ponderomotive force that till certain moment are capable of compensating the Archimedes force causing upward flow of the water. In addition to the described delay of natural convection the magnified images of water exposed to high-power laser beam revealed another important result – formation of a long channel with rather well defined boundary that has slightly diverging geometric outline. This channel is coaxial with the laser beam.

There are several volumetric forces acting on the liquid that can cause convection during laser interaction. These forces can be divided into two classes: thermally induced forces and ponderomotive forces due to the gradient of electric field. The former includes Archimedes force induced due to the liquid heating, evaporation recoil induced force [15,16] and Marangoni convection [16,17] and the latter includes electrtostictive force and photon pressure induced force. To the best of our knowledge our research is the first in which ponderomotive forces are considered in the context of laser effect on water. All prior works concentrated attention mostly on thermal effects induced during laser interaction, such as evaporation recoil induced flow and Marangoni convection [15-17]. Comparison of simple estimates with the computation results that are presented below can clearly show that, under the considered laser interaction conditions, both evaporation recoil and Marangoni convection have negligible effect due to the relatively low water temperatures and small temperature gradients in the near surface area. Thus, the dominant forces under the laser interaction conditions similar to the described above are the ponderomotive and Archimedes forces. The amplitude of these forces is determined by the nonuniformity of the laser beam intensity or, more precisely, by the gradient of the amplitude of electric fiels in the laser beam. As we will show below, under certain laser intraction conditions the ponderomotive force induced pressure generates convection that is dominant at he initial stages of interactioin producing convective cooling of the water chanel heated due to the laser absorption. This ponderomotive (electrostriction + photon pressure) indiced convection also interferes with the thermally induced convection driven by the Archimedes force.

It is important to note here that authors of works [18-20] considered physical effects that occur on the boundary between condensed and gaseous phases where the "sharp jump" of the properties and, in particular, sharp change of the coefficient of refraction, takes place. Unlike these works that considered surface absorption of laser radiation (i.e. infinitely small absorption length), our work considers volumetric effects caused by the absorption of laser radiation within liquid while propagating relatively large distance (i.e. finite and relatively large absorption length). We show in our theoretical consideration supported by the experimental evidences that under such conditions water acquires momentum from the laser radiation due to such volumetric absorption and this effect is different from the photon pressure.

## II. PHYSICAL MODEL AND EQUATIONS

Here we assume laser parameters same as in our previous experient, i.e. CW laser with wavelength $\lambda$=1.064μm and characteristic radius of the laser beam $r_b$=3mm. We also disregard beam convergence and assume cylindrical beam. This assumption is justified because the distances of beam propagation that will be considered are smaller than 1 cm and the radius of the beam decreases by only 20% as the beam propagates from the surface of the water to the depth 1 cm. In addition, the taken shlieren video indicates possible beam defocusing such that it compensates geomertic convergence producing nearly cylindrical channel.

Lest us first estimate characteristic time of conduction cooling. The thermal diffusion coefficient of water is $\chi_T = k/\rho c_p$, where $k \approx 0.6$ W/m·K is the heat conductivity; $c_p \approx 4200$ J/kg·K is the heat capacity. The estimate gives the following value for thermal diffusion of water: $\chi_T \approx 1.4 \cdot 10^{-7}$ m$^2$/s. The characteristic time for heat-conduction cooling of a cylindrical heating region of a radius $r_b$ is $\tau_T \approx r_b^2/\chi_T$. Thus, for a channel with a radius $r_b = 3$ mm the characteristic cooling time is approximately $\tau_T \approx 60$ sec. This time exceeds considerably the characteristic times that we considered in our simulations, and, the conduction cooling starts playing role at the times significantly larger than the time of convection development observed in our experiments (Figures 1 and 2). Therefore, further we will not take into consideration the heat losses due to the thermal conductivity.

Now, let us describe the ponderomotive hydrodynamic effect. In general, the volumetric force acting on the dielectric fluid in non-uniform electric field is determined by the Helmholtz equation [4-6], which in the case of a quasineutral homogenious polar dielectric fluid immersed in a nonuniform electromagnetic field can be expressed as [6,11]

$$\vec{F}_p = \frac{\varepsilon_0}{2}\nabla\left(E^2 \frac{\partial \varepsilon}{\partial \rho}\rho\right) \approx \frac{\varepsilon_0(\varepsilon - 1)(\varepsilon + 2)}{6}\nabla E^2, \qquad (1)$$

Taking into account that at the optical frequencies water can be considered as a non-polar liquid and, therefore, with dielectric constant and refractive index related as $\varepsilon \approx n^2$, the irradiance in the laser beam is given by the following formula:

$$I = \frac{\varepsilon_0 \varepsilon E_a^2}{2} \frac{c}{n} \approx \frac{\varepsilon_0 n c E_a^2}{2} = \varepsilon_0 n c E^2, \tag{2}$$

where $E = \frac{E_a}{\sqrt{2}}$, $E_a$ are the electric field rms and amplitude values, $n$ is the real part of the refractive index (for laser wave length in a near IR range, $n \approx 1.33$), and $c$ is the speed of light.

Assuming Gaussian radial distribution of irradiance and disregarding beam divergence, the irradiance can be expressed as follows

$$I(r,z) = I_0 e^{-r^2/r_b^2 - \mu z}, \tag{3}$$

where $I_0 = \frac{P_L}{\pi r_b^2}$ is the maximal irradiance at the water-air interface $(z = 0)$, $P_L$ is the laser power, and $\mu$ is absorption coefficient. Then, from equation (1) using equations (2,3) we find radial and axial components of electrostrictive force acting on water

$$F_{p,r}(r,z) = \frac{(n^2-1)(n^2+2)}{6cn} \frac{\partial I}{\partial r} = -\frac{(n^2-1)(n^2+2) r I(r,z)}{3cnr_b^2} \tag{4}$$

$$F_{p,z}(r,z) = \frac{(n^2-1)(n^2+2)}{6cn} \frac{\partial I}{\partial z} = -\frac{(n^2-1)(n^2+2) \mu I(r,z)}{6cn} \tag{5}$$

The force components given by equations (4,5) are the corresponding components of the negative gradient of electrostriction pressure:

$$p_E(r,z) = -\frac{(n^2-1)(n^2+2) I(r,z)}{6cn}. \tag{6}$$

Due to laser beam absorption, water gains momentum in the direction of the beam propagation and, therefore, water is subjected to additional volumetric force – photon pressure force, $\vec{F}_{opt}$, with radial and axial components given by

$$F_{opt,r} = 0; \quad F_{opt,z} = -\frac{n}{c} \frac{\partial I}{\partial z} = \frac{n \mu I}{c}. \tag{7}$$

Now we consider two-dimensional axysimmetrical nonstationary problem for a viscous incompressible fluid ($\rho = \text{const}$, where $\rho$ is the fluid density) [18] in gravitational field and exposed to a laser beam:

$$\nabla \cdot \vec{u} = 0, \tag{8}$$

$$\rho\left(\frac{\partial \vec{u}}{\partial t}+(\vec{u}\cdot\nabla)\vec{u}\right)=-\nabla(p+p_E)+\rho\vec{g}\big(1-\beta(T-T_0)\big)+\vec{F}_{opt}+\eta\Delta\vec{u}\,, \tag{9}$$

$$\rho c_p\left(\frac{\partial T}{\partial t}+(\vec{u}\cdot\nabla)T\right)=Q+\lambda\Delta T+\frac{\eta}{2}\left(\frac{\partial u_i}{\partial x_k}+\frac{\partial u_k}{\partial x_i}\right)^2, \tag{10}$$

where $Q(r,z)=\left|\frac{\partial I}{\partial z}\right|=\mu I$ is the heat source due to laser beam absorption, the absorption coefficient in water at the considered wavelegth of Nd:YAG laser is $\mu \approx 16\,\mathrm{m}^{-1}$ [21], $\eta$ is the dynamics viscosity of water and $\eta(T=293K)\approx 1.05\cdot 10^{-3}\ \mathrm{Pa\cdot s}$, $\vec{g}$ is the acceleration due to gravity, and $\beta$ is the coefficient of volumetric thermal expansion which is $\beta(T)=-\frac{1}{\rho}\left(\frac{\partial\rho}{\partial T}\right)$ [17].

Static pressure in water can be found by solving Poisson equation for instantanous distributions of velocities and forces obtaind from equation (9) as a result of applying operator $\nabla$ to both sides of this equation:

$$\Delta p=-\rho\left[\left(\frac{\partial u_z}{\partial z}\right)^2+2\frac{\partial u_r}{\partial z}\frac{\partial u_z}{\partial r}+\left(\frac{\partial u_r}{\partial r}\right)^2+\left(\frac{u_r}{r}\right)^2\right]-\Delta p_E+\nabla\left[\vec{F}_{opt}-\vec{g}\beta(T-T_0)\right] \tag{11}$$

The system of equations (8) is solved using standard set of boundary conditions at applied at the limits of the computational area $z=0,h$ and $r=0,R$ for pressure

$$\begin{aligned}
&p(z,R)=p_0+\rho g z\\
&\frac{\partial p(z,0)}{\partial r}=0\\
&\frac{\partial p(h,r)}{\partial z}=\rho g\\
&p(0,r)=p_0+\frac{I(0,z)}{2c}\left[\frac{(n^2-1)(n^2+2)}{3n}-(n-1)\right]
\end{aligned}\quad, \tag{12}$$

and for velocity components

$$\frac{\partial u_{z,r}(z,R)}{\partial r}=0$$
$$\frac{\partial u_{z,r}(h,r)}{\partial z}=0$$
$$\frac{\partial u_{z,r}(z,0)}{\partial r}=0 \quad . \tag{13}$$
$$\frac{\partial u_{r}(0,r)}{\partial z}=0$$
$$u_z(0,r)=0$$

The last equation in the set of equations (12) that gives boundary condition for pressure on the boundary between dielectric and air [4] is obtained from the conservation of flux of electro-magnetic energy averaged during a period of the electro-magnetic wave and neglecting reflection on the boundary. The latter is valid approximation since the water reflectivity at the laser wavelength λ=1064nm is small. Indeed, using formula for the reflectivity at the boundary of absorbing material [20] we have for water $R=\frac{(n-1)^2+k^2}{(n+1)^2+k^2}\approx 0.02$, where $n$=1.33 is the real part of the refractive index of water [22], $k=\frac{\mu\lambda}{4\pi}\sim 10^{-6}$ is the imaginary part of refractive index, and $\mu\approx 16\,\mathrm{m}^{-1}$ is the absorption coefficient of water [19]; all values are for the Nd:YAG laser wavelength.

Additionally, in our consideration we assume negligible effect related to the change of water surface geometry produced during laser-water interaction. Thus, we disregard the effects of Laplace pressure determined by surface tension and water surface curvature and evaporation recoil. It is justified by comparing considered laser interactions conditions to those that produce significant effect.

## III. RESULTS AND DISCUSSION

The pressure difference caused by the ponderomotive effect forces water toward laser beam axis as a result of electrostriction and, because water is practically incomressibe and photon pressure effects, axial water flow develops moving water from the area of higher irradiance toward the area of lower irradiance. Simultaneously absorption of laser beam in water causes increase of water temperature. Initailly, when temperature increase is insignificant the ponderomotive induced convection dominates; however, as water temperature rises the Archimedes force increases resulting in increase of the effect and domination of ordinary thermal convection.

The Archimedes force dependence on temperature is

$$F_A=g\rho\beta(T)(T-T_0)\,. \tag{14}$$

Thus, when $F_A>F_{p,z}+F_{opt,z}$, the thermal convection dominates over the ponderomotive induced convection. This transition takes place when temperature increase becomes larger than

$$\Delta T = T - T_0 > \frac{I(r,z)\mu(5n^2 - n^4 + 2)}{6nc\rho g\beta}. \quad (15)$$

For example, for laser power $P_L = 150\,\mathrm{W}$ the thermally induced convection dominates ponderomotive convection at the beam axis near the water surface when increase of temperature exceeds approximately $\Delta T \approx 0.21\,\mathrm{K}$. This estimate, made for $T \approx T_0 = 293\mathrm{K}$ and $\beta(T) \approx 1.82 \cdot 10^{-4}\ \mathrm{K}^{-1}$[23], is close to the results of numerical computations shown in the Figure 4. The evolution of water temperature and relative change of density, $\delta\rho / \rho = \beta(T)(T - T_0)$, at the laser beam axis are shown in the Figure 5. The positive velocity corresponds to the motion in the direction of laser beam propagation (ponderomotive convection dominant). As the computations show that with time the temperature of water increases and the ordinary thermal convection becomes dominant and the sign of water flow velocity changes to negative indicating that heated water raises due to higher buoyancy.

It is curious to consider that in the environment of zero- or microgravity the ponderomotive convection will take place unopposed by the thermal convection. Under such conditions the fluid velocity can reach significant values in just a few seconds. This effect can be practically utilized for contactless mixing of fluids and solutions in space labs. The example of computation for zero-gravity and as above laser parameters is shown in the Figure 6.

Another intriguing case is when in the gravity environment the beam of laser is directed upwards. In this case the direction of ponderomotive convection flow coincides with the direction of thermally induced convection and the liquid flow velocity will increase up to the value determined by viscosity.

It is worth mentioning that laser fluid interaction can be dramatically different in case of large laser beam absorption. For example, for $CO_2$ laser with the wavelength $\lambda_L = 10.6\,\mu\mathrm{m}$, water is highly absorbing with absorption coefficient $\mu \approx 7 \cdot 10^4\,\mathrm{m}^{-1}$ [21]. In this case, when most of the laser power is absorbed in the thin near surface layer, $\Delta z \sim \mu^{-1} \approx 14.3\,\mu\mathrm{m}$, neither ponderomotive nor thermal convection has time to develop. Relatively quickly water reaches high temperatures even for relatively small laser powers and intense surface evaporation occurs [14] generating substantial recoil force that produces fast liquid ejection and formation of vapor channel in the liquid known as "keyhole" [15,16]. This is topic of significant research effort and it is beyond the scope of our work.

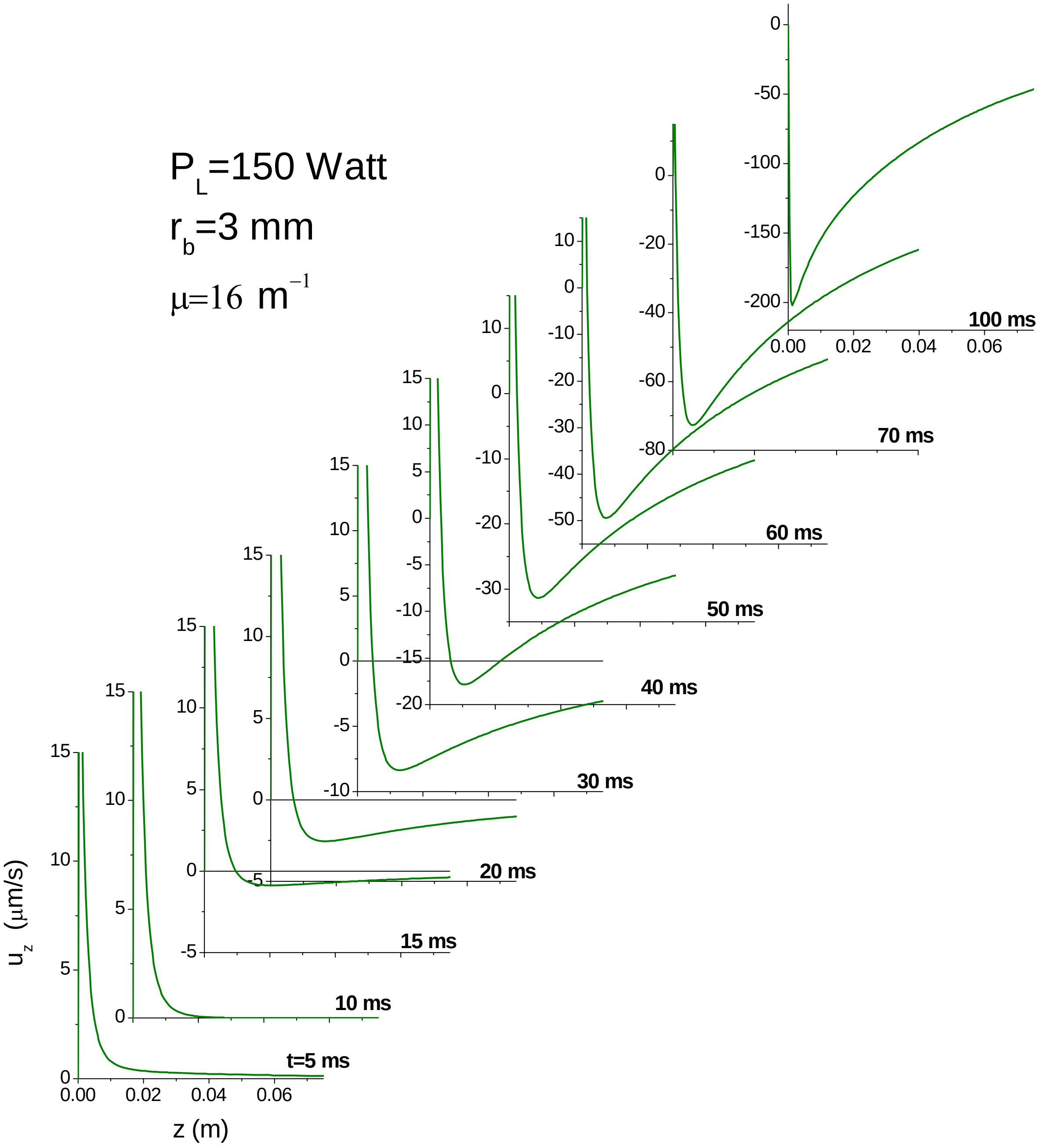

Fig.4. Velocity of cw Nd:YAG laser induced convection in water at the beam axis (z-axis). The surface of water is at z=0 and the z-axis is normal to the water surface and directed down, i.e. along the vector of gravitational acceleration. Time is counted from the moment of turning laser on. Positive velocity direction is in the z-axis direction.

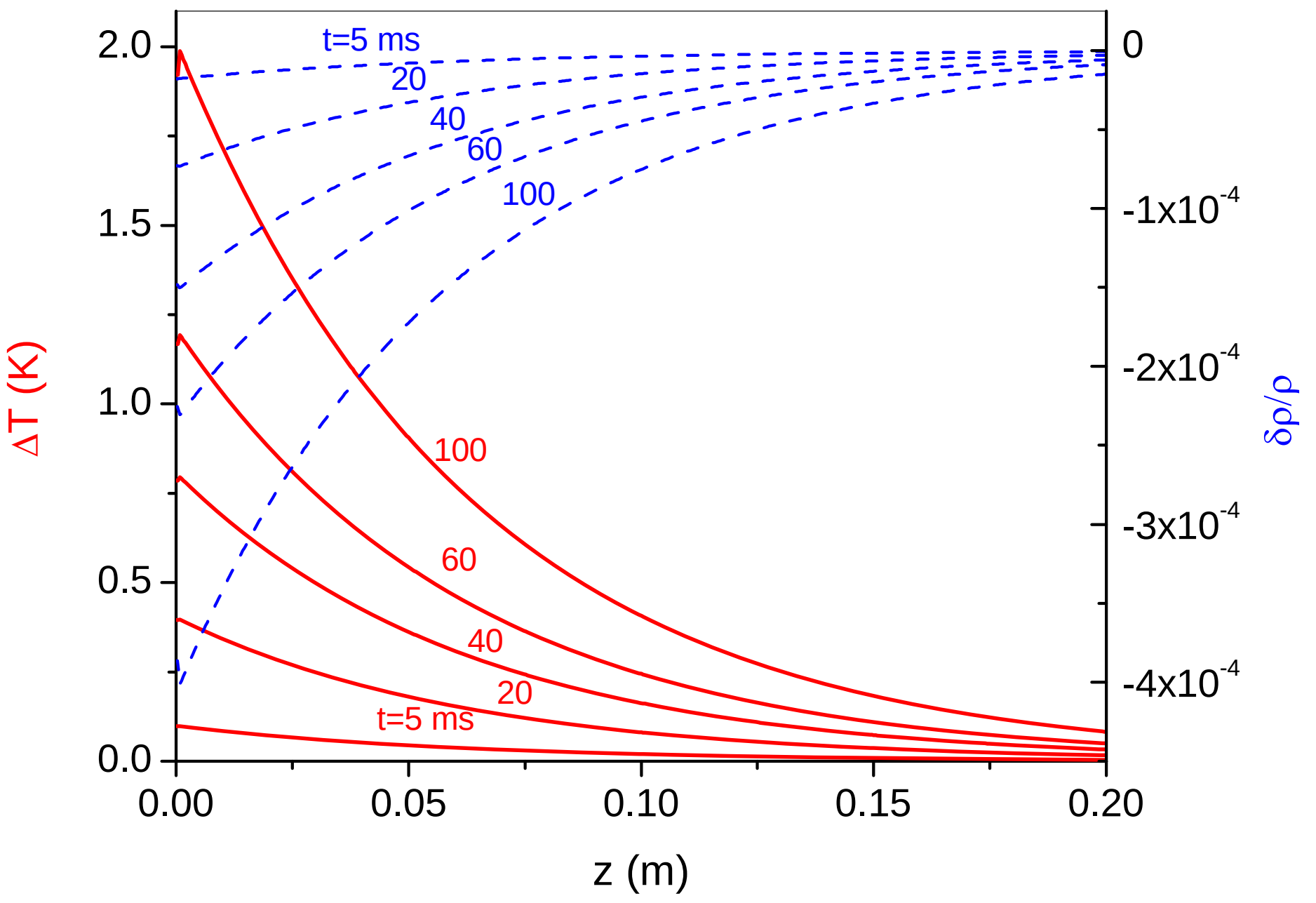

Fig.5. Distributions of temperature change and relative density of water along the beam axis for different moments of time. Power of Nd:YAG cw laser is 150W and the beginning of laser irradiation as at t=0.

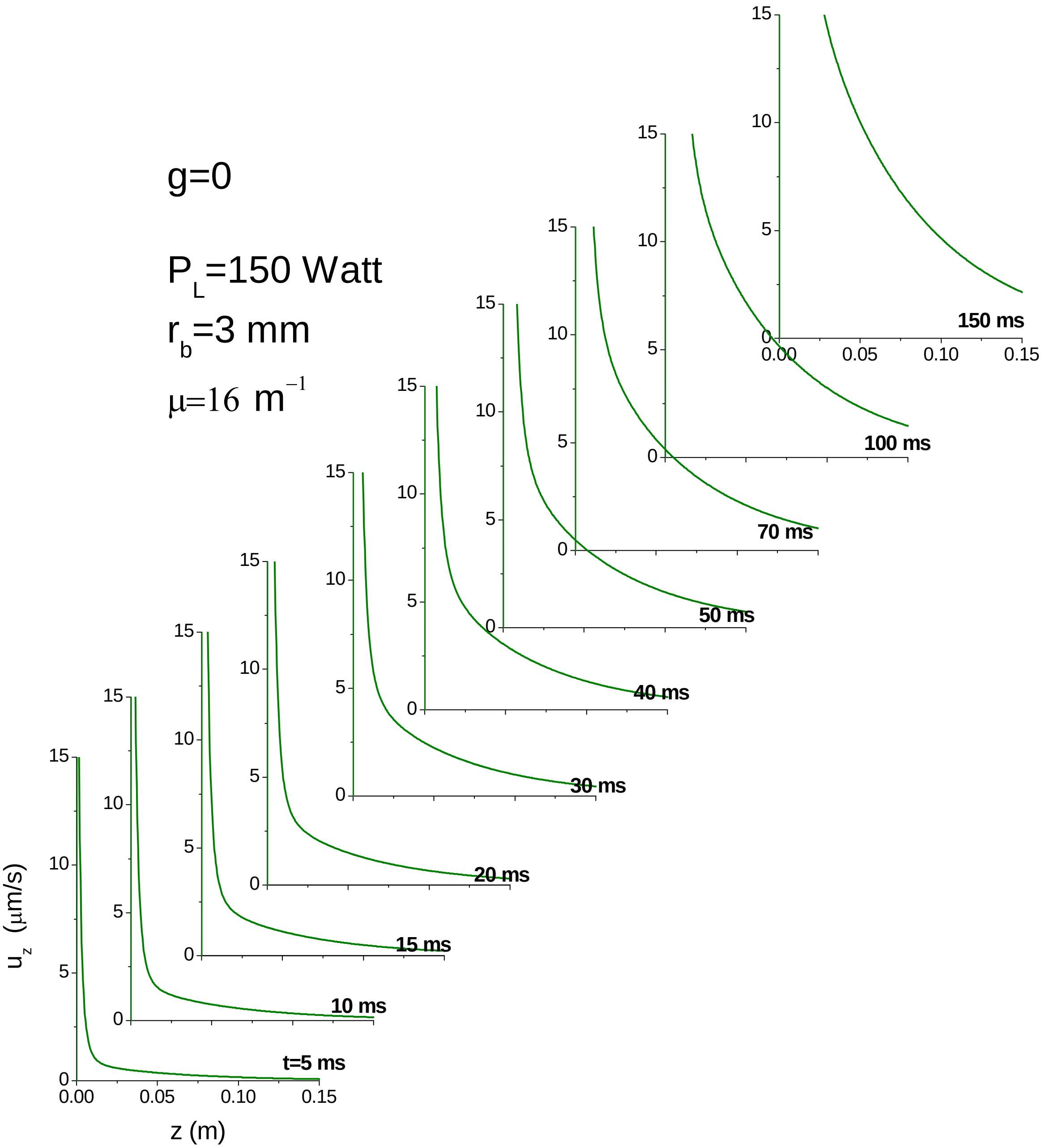

Fig.6. Velocity of cw Nd:YAG laser induced convection in water at the beam axis (z-axis) in zero gravity. The surface of water is at z=0 and the z-axis is normal to the water surface and directed down, i.e. along the vector of gravitational acceleration. Time is counted from the beginning of laser irradiation. Velocity of the flow is positive (in the direction of laser beam propagation) since ponderomotive convection is unopposed by thermal convection.

For the sake of demonstration of the effect of laser beam power on the ponderomotive convection the temporal evolution of water temperature and longitudinal component of flow velocity computed for different locations along the laser beam are shown in the Figure 7 for laser power of 150 W (a-c) and 1500 W (d-f). Same as above, the positive values of velocity at the beginning of exposure correspond to the downward motion along the beam propagation when the ponderomotive force is dominant and the negative velocity corresponds to the upward flow when Archimedes force driven convection is dominant. The computations show that regular thermal convection develops with noticeable delay of ~10 ms. The local increase of water temperature when the thermal convection overtakes the ponderomotive convection is close to the estimate given by equation (15). At later stages the convective flow is faster for higher laser power and the convection velocity reaches values from several mm/s to several cm/s in several tenths of a second, same as observed in the experiments.

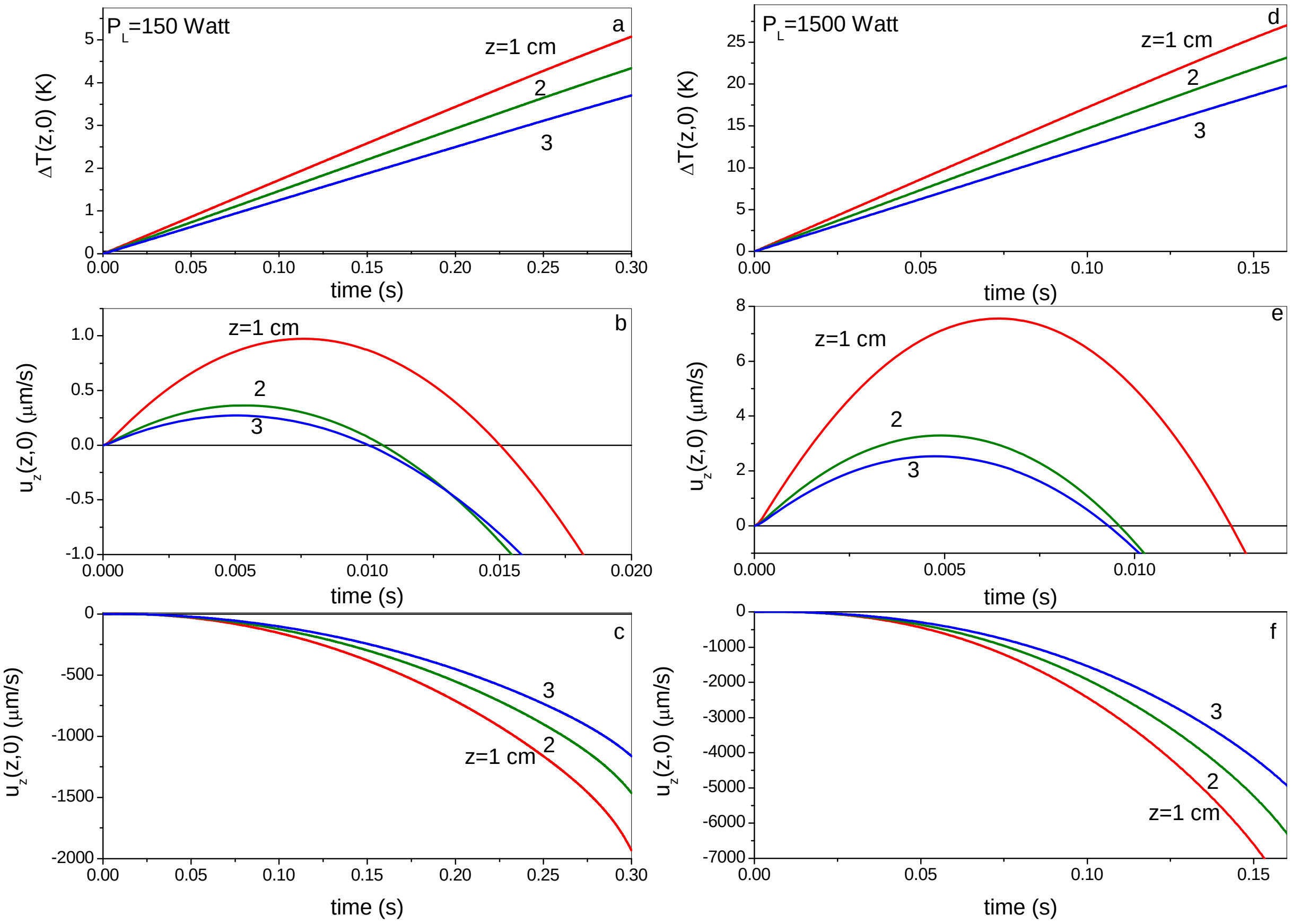

Fig.7. Temporal change of water temperature increase and flow velocity in different locations at the beam axis for laser power 150W (a-c) and 1500W (d-f). The surface of water is at z=0 and the z-axis is normal to the water surface and directed down, i.e. along the vector of gravitational acceleration. Time is counted from the beginning of laser irradiation. Positive velocity direction is in the z-axis direction.

Additionally, the spatial distribution of water temperature and longitudinal component of flow velocity along the beam axis computed for laser power of 150 W and 1500 are shown in the Figure 8 after 5ms (a,b) and after 150ms (c,d) from the beginning of exposure. One can see that at the instance of 5 ms the flow velocity at the beam axis is directed down into the liquid. At later stages, as for 150ms, the high velocity thermally driven upward directed convection develops.

The velocity of this “regular” convection increases until becoming unstable and resulting in efficient turbulent mixing and cooling of water. Our model does not include turbulence instability of the water flow; however, to some degree, this transition to turbulent flow is reflected in the loss of computational stability of our code when increase of water temperature approaches 30-40 degrees.

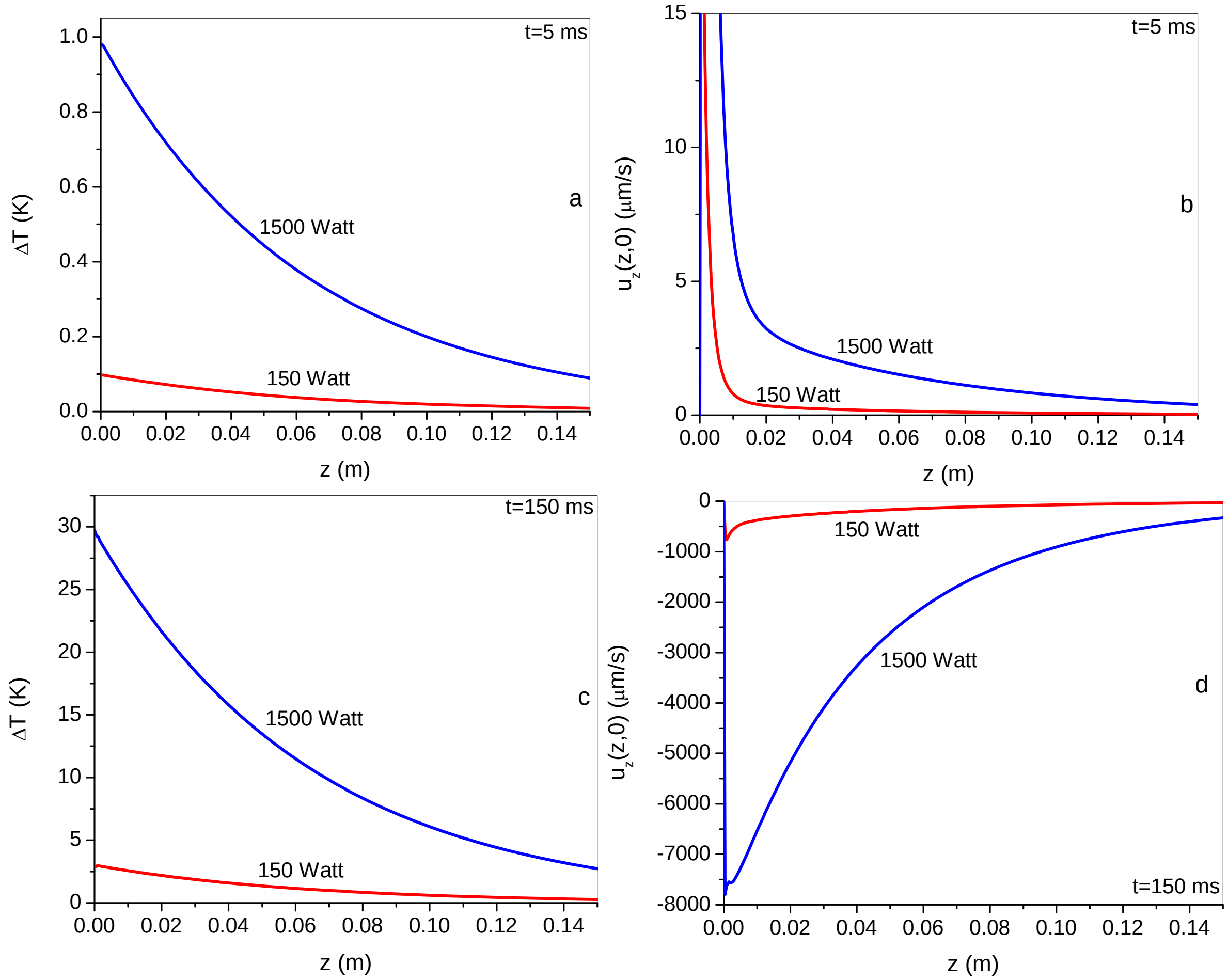

Fig. 8. Distributions of temperature change and flow velocity of water along the beam axis computed for 150 W and 1500W of laser power in different moments of time from the beginning of laser irradiation: (a),(b) – 5 ms and (c), (d) – 150 ms.

## IV. CONCLUSIONS

We have created a theoretical model that describes macroscopic fluid motion induced by the optical pressure and ponderomotive effect that takes place during interaction of laser beam with absorbing liquid. This numerical simulations performed using this model showed that ponderomotive convection produces flow in the direction of laser beam propagation. The simulation results also showed that water is heated due to the absorption of laser beam and, as a result of water temperature increase, the well-known thermal convection is induced by Archimedes force. The characteristic time of development of ponderomotive convection is shorter than the time during which thermal convection is established. Because the ponderomotive induced and thermally induced convective flows have opposite directions the turbulent water mixing develops when flow induced by the Archimedes force overtakes the ponderomotive force induced flow.

Fast development of water flow driven by the optical pressure and ponderomotive force produces initial cooling preventing rapid increase of temperature and water evaporation that is expected from the estimate that doesn't include ponderomotive convection. The thermal convection that develops on later times further contributes to water cooling and suppression of intense evaporation since turbulent mixing is produced by counter flows driven by laser induced and Archimedes forces.

## ACKNOWLEDGEMENTS

The authors express their deep gratitude to Dr. M. Wardlaw of the ONR for sharing our fascination with physics of water and for the ONR for the funding support that made this research possible (Award: N00014-15-1-2109). We would also like to thank staff of the Pennsylvania State University ARL Laser Division where the experiments were conducted.